# The Making of Digital Ghosts: Designing Ethical AI Afterlives

Giovanni Spitale[1,*] 0000-0002-6812-0979

Federico Germani[1,] 0000-0002-5604-0437

1: ITE Lab, Institute of Biomedical Ethics and History of Medicine, University of Zurich, Zurich, Switzerland.

*: Corresponding Author. giovanni.spitale@ibme.uzh.ch

## Abstract

Advances in artificial intelligence now make it possible to simulate the dead through chatbots, voice clones, and video avatars trained on a person's digital traces. These "digital ghosts" are moving from fiction to commercial reality, reshaping how people mourn and remember. This paper offers a conceptual and ethical analysis of AI-mediated digital afterlives. We define what counts as a digital ghost, trace their rise across personal, commercial, and institutional contexts, and identify core ethical tensions around grief and well-being, truthfulness and deception, consent and posthumous privacy, dignity and misrepresentation, and the commercialization of mourning. To analyze these challenges, we propose a nine-dimensional taxonomy of digital afterlife technologies and, building on it, outline the features of an ethically acceptable digital ghost: premortem intent, mutual consent, transparent and limited data use, clear disclosure, restricted purposes and access, family or estate stewardship, and minimal behavioral agency. We argue for targeted regulation and professional guidelines to ensure that digital ghosts can aid remembrance without slipping into forms of deception.

## Keywords

digital ghosts; deadbots; griefbots; digital resurrection; privacy; digital remains; residual personhood; postmortem data; autonomy; dignity; mourning; AI ethics; death and technology; deception

# Introduction

The rise of artificial intelligence has transformed not only how we live, but also how we grieve. Over the past decade, a growing number of technologies, ranging from chatbots trained on personal data to video avatars and voice clones, have emerged to simulate the presence of the deceased (Hollanek and Nowaczyk-Basińska 2024; Klugman 2024). These so-called "deadbots" or "digital ghosts" allow users to converse with AI-generated representations of loved ones who have passed away, offering a new and deeply intimate form of digital memorialization. While some view these tools as comforting extensions of memory, others see them as unsettling illusions that blur the boundaries between life, death, and simulation.

This paper provides a conceptual analysis of the emerging phenomenon of AI-driven posthumous representations, focusing on their ethical implications and real-world applications. Drawing from documented cases, such as Joshua Barbeau's use of Project December to recreate his late fiancée, and the 2025 Arizona courtroom appearance of an AI-generated avatar of murder victim Christopher Pelkey, we explore how digital ghosts are reshaping experiences of mourning, memory, and identity.

In what follows, we first define what counts as a "digital ghost" and trace the rapid rise of posthumous AI simulations across personal, commercial, and institutional contexts. We then examine the ethical tensions these technologies create around grief, truthfulness, consent, privacy, dignity, and commercialization; before offering a nine-dimensional taxonomy that clarifies the spectrum of digital afterlife systems now emerging. Building on this taxonomy, we articulate the features that an ethically acceptable digital ghost should possess, and identify the deeper normative commitments that ground these judgments. We conclude by reflecting on how digital ghosts challenge cultural understandings of mourning, memory, and the boundaries between life and death.

## Defining digital ghosts and "deadbots"

Digital ghosts, also known as deadbots, griefbots, generative ghosts, death avatars, or postmortem avatars, are AI-driven digital representations of deceased individuals (Hollanek and Nowaczyk-Basińska 2024; Klugman 2024; Morris and Brubaker 2025; Hurshman et al. 2025). These systems (often powered by large language models) are trained on the digital footprints people leave behind – their social media posts, emails, text messages, photos, videos, voice recordings, and other data (Hollanek and Nowaczyk-Basińska 2024; Klugman 2024; Morris and Brubaker 2025; Hurshman et al. 2025). The result is an interactive simulacrum that can converse in the first person as if it were the departed, mimicking their speech patterns, personality, and even visual likeness. In other words, a digital ghost uses a dead person's data to respond like that person; an idea so plausible that in 2021 Microsoft was even granted a patent for an individualized chatbot capable of doing exactly this (Lindemann 2022; Brown 2021; Abramson and JR 2020).

Digital ghosts are software constructs that come in various forms: some are purely text-based chatbots, others can speak in the person's voice, and more advanced versions appear as video avatars (for example, simulating a Zoom call with the deceased)(Klugman 2024; Hollanek and Nowaczyk-Basińska 2024). The cost and complexity can vary widely, from simple text chat services that charge a few USD per session, to sophisticated custom video avatars costing thousands of USD (Klugman 2024; Hollanek and Nowaczyk-Basińska 2024). Regardless of form, all digital ghosts share the core feature of offering a way to interact with a representation of someone who has died, effectively creating an illusion (or perhaps extension) of life after death in the digital realm (Morris and Brubaker 2025; Hurshman et al. 2025).

## The rise of digital ghosts

Once again, developments long imagined in science-fiction narratives – ranging from *Star Trek* (Star Trek: Discovery 2024; Star Trek: The Next Generation 1993) to Black Mirror (Black Mirror 2013) – have moved into the realm of technological reality. Scenarios that were once treated as speculative fiction, such as the AI-mediated resurrection depicted in the Black Mirror episode 'Be Right Back,' have rapidly transitioned into real possibilities. In the mid-2010s, early prototypes of griefbots appeared. A notable example in 2016 was Eugenia Kuyda's creation of a chatbot from her friend Roman Mazurenko's text messages after he died in an accident (The Guardian 2016; Huet 2016). Kuyda's "Roman bot" gained media attention as one of the first real-world instances of talking to a dead loved one via AI; she argued that "it's not about pretending someone's alive. It's about accepting it…and not staying in denial" (The Guardian 2016) positioning the technology as a tool for mourning, rather than make-believe. Allegedly, the codebase of "Roman Bot" became the foundation for Replika, a popular companion bot used as a "friend", "partner", or "mentor" – in fact mostly for sexting (Huet 2023; 2016). Around the same time, others pursued similar ideas: for example, journalist James Vlahos built a "Dadbot" to preserve his terminally ill father's stories and personality, which later evolved into a service called HereAfter AI for creating "life story avatars." By the late 2010s, startups like Eternime were exploring the "digital afterlife" business model, and the concept of "digital immortality" was entering public discourse.

The period 2020–2021 marked a turning point that thrust digital ghosts into wider awareness. In 2021, Project December – an online platform that let users create chatbots of anyone, real or fictional – made headlines when a young man named Joshua Barbeau used it to chat with an AI simulation of his fiancée eight years after she had passed away (Hollanek and Nowaczyk-Basińska 2024). This emotional story (extensively reported by the San Francisco Chronicle) captured the public imagination and highlighted both the allure and the unease of such technology (Fagone 2021). In Barbeau's case, the chatbot's eerily lifelike responses provided comfort and a sense of farewell, but the situation also raised red flags for the tech's creators – OpenAI temporarily cut off Project December's access to its GPT-3 language model, citing concerns about creating bots that mimic real people without explicit consent (Hollanek and Nowaczyk-Basińska 2024). The Jessica Simulation case thus became a cautionary tale for industry guidelines, showing how quickly AI grieving tools had outpaced ethical frameworks.

By the mid-2020s, digital resurrection had become a growing industry. Numerous services emerged to help people preserve or recreate loved ones: for example, California-based StoryFile offers interactive pre-recorded video interviews (so one can "interact" with a deceased person's video), and HereAfter AI provides audio chat experiences based on a person's recorded stories (Grieshaber and Hadero 2024). New startups continue to appear – Eternos recently announced it had created a "comprehensive, interactive AI version" of a terminally ill man to comfort his family after his impending death (Grieshaber and Hadero 2024). There are even tongue-in-cheek offerings like "Seance AI," which simulates a faux supernatural séance via chatbot. Meanwhile, Project December reinvented itself with its own models and still invites users to "simulate the dead" for a small fee (Grieshaber and Hadero 2024).

The phenomenon is global. In China, where ancestor veneration is culturally significant, companies have embraced AI re-creations. By 2022–2024, Chinese startups were offering basic "grief avatar" services for as little as ¥20 (about 3 dollars) – allowing mourners to upload a photo and get a moving, speaking digital avatar of their loved one (Hawkins 2024). More sophisticated projects also made news: in 2023, the tech company SenseTime even unveiled a digital clone of its founder, Tang Xiao'ou, to deliver a speech months after he had died, by training an AI on his voice and video recordings (Hawkins 2024). This dramatic demonstration underscored how far the tech has advanced in realism. Chinese developers report a surge in demand: one Weibo engineer said he had "helped more than 600 families 'achieve reunion' with their loved ones" for the annual Tomb-Sweeping Festival (Hawkins 2024). However, not all such resurrections are welcomed – for instance, when fans created an unauthorized AI avatar of the late singer Qiao Renliang, his parents were outraged, saying the video "exposed scars" and was made without their consent (Hawkins 2024). This incident highlighted the sensitivities around consent and the potential for harm when third parties resurrect someone's likeness against the family's wishes.

In addition to personal and family use, digital ghosts have seen novel hybrid and public applications. In 2025, journalist Jim Acosta publicly interviewed an AI recreation of Joaquin Oliver, a student killed in the 2018 Parkland shooting. Created by Oliver's parents as part of their gun-control activism – and initially used privately to hear their son's voice again – the digital ghost appeared on national television, blurring the line between intimate mourning practices and public advocacy (Jim Acosta 2025; Hurshman et al. 2025). Such cases illustrate how digital ghosts are becoming actors in civic and political arenas, not merely companions for the bereaved.

In this sense, a landmark U.S. case pushed the border even further. In 2025, an Arizona court allowed a murder victim to "speak" at his own sentencing via an AI avatar. The family of Chris Pelkey, who had been killed in a road-rage incident, used AI to create a video of Pelkey delivering a posthumous victim impact statement (Duffy 2025). In what's believed to be a legal first, the lifelike avatar addressed the courtroom (and the shooter) with forgiving words, fulfilling the family's wish that "he [speak] for himself." The presiding judge even remarked that he "loved that AI", finding it a powerful way to convey the victim's perspective. This extraordinary example shows how digital ghosts are not only used in private grieving but are crossing into domains like justice – raising questions about how far we should go in letting simulations of the departed participate in the affairs of the living.

## Ethical tensions

These emerging practices have outpaced the ethical and legal frameworks needed to govern them, creating a series of tensions that now demand systematic scrutiny. The advent of digital ghosts creates a profound tension between memory and deception. On the one hand, these tools offer a new way to remember and remain connected with those we have lost; on the other hand, they involve engaging with something we know is an artificial construct. This tension raises difficult questions: Is it healthy or harmful to prolong contact with a deceased loved one through a simulation? Is it a kind of therapeutic "memory" or a dangerous self-deception?

Do griefbots help heal, or do they hinder the natural grieving process? The answer may depend on the individual and how the technology is used. It has been proposed that interactive AI memorials can comfort the bereaved by providing a gradual, controlled way to say goodbye. In line, according to previous work, nearly 96% of people naturally continue bonds with the dead through things like dreams or talking to a gravesite (Klugman 2006). In that light, a griefbot could simply be another medium for an

"ongoing bond," potentially easing the pain in "small batches" over time (Klugman 2024). Some early users have reported that talking to a deceased loved one's avatar felt cathartic and helped them cope with loneliness or guilt, rather than denying the death (Lindemann 2022; Krueger and Osler 2022). Critiques, however, focus on the fact that reliance on a deadbot might reinforce denial or prolong grief rather than resolve it, and that people could become dependent on these tools, finding it so hard to disconnect that they get "locked" in a state of mourning and complicated grief (Voinea et al. 2025; Klugman 2024; Lindemann 2022). If one spends hours every day chatting with a late spouse's simulacrum, is it truly helping to heal, or just reopening the wound repeatedly? Empirical research is still scarce, therefore Lindemann argues that griefbots should be treated as experimental therapeutic devices – their safety and efficacy ought to be proven before they are widely used (Lindemann 2022).

Arguably using an AI to emulate a person raises the issue of truthfulness. The user knows intellectually that they are not really speaking with the deceased person, yet they may willingly suspend disbelief to derive comfort from the interaction (Lindemann 2022; Voinea et al. 2025; Krueger and Osler 2022). Is this a benign illusion, akin to imagining what the dead person would say if they were here? (Klass et al. 1996) Or is it fundamentally an unethical deception – a lie we tell ourselves that prevents acceptance of reality? A key factor is whether the mourner remains aware of the bot's artificial nature. If so, engaging with the digital ghost might be viewed as a conscious coping ritual rather than genuine confusion. In practice, talking to a bot could be comparable to writing letters to the deceased or reminiscing out loud – activities many find healing (Stroebe and Schut 1999; Neimeyer 2019). However, the line can blur, especially for more vulnerable users, such as young children. There is a risk of a "blind spot" where one starts to treat the AI's responses as if they carry the authority or spirit of the actual person, rather than mere algorithmic predictions (Mamak 2024). This becomes problematic if, say, someone starts making life decisions "because my AI mom said so," or genuinely believes the deceased's consciousness might reside in the program. Overall, the balance here pits compassion against honesty: is it ethical to create a knowingly false persona if it soothes someone's pain? Some have argued that comparable acts of deception can be morally permissible (Isaac and Bridewell 2017; Sharkey and Sharkey 2021), while others worry it amounts to indulging a comforting falsehood (Danaher 2020; Kaczmarek 2025).

One of the thorniest issues is who is to be digitally resurrected – and who gets to decide. Most deceased individuals never explicitly consented to having a digital ghost made from their personal data. Does using their emails, social posts, videos, or other data in this way violate their posthumous privacy or dignity? In life, people often have rights over their personal data, but legally those rights usually expire at death – for example, Europe's GDPR recital 27 explicitly excludes from protection the data of the deceased (EU GDPR 2016). Ethically, however, we might feel that a person's "digital remains" deserve respect much like their physical remains do (UN General Assembly 2024; Birnhack and Morse 2022; Richardson 2015). If a loved one's data is locked behind passwords, many jurisdictions require a court order or executor's permission to access it – indicating an intuition that not anyone can just appropriate it. Yet currently griefbot services often operate in a grey area, scraping publicly available data or letting users upload chat logs they possess, without any clear legal guidelines. This raises the question: who owns and controls your digital footprint after you die? (Lim 2024). Is it your family (heirs), the platforms that store it (big tech companies), or does it effectively fall into the public domain? Scholars note a "lack of clear prohibitions" against people inputting others' data to create griefbots (Lim 2024), which means by default it's happening without robust oversight. Hollanek and Nowaczyk-Basińska argue that at minimum, informed consent should be required – ideally the explicit consent of the person while alive, or at least the permission of their next-of-kin – before a deadbot is created (Hollanek and Nowaczyk-Basińska 2024). They also propose a principle of "mutual consent": the deceased (as data donor) must have agreed, and the user (survivor) must also consent to engage (Hollanek and Nowaczyk-Basińska 2024). Without such consent, we risk infringing on what little posthumous privacy and autonomy a person can have. After all, a digital ghost might reveal things the person never chose to share publicly. Even if a person did leave data behind, they might not have intended it to be repurposed in a new interactive form. The intent matters – for instance, someone might be fine with their diary being read by family after death, but not with an AI using that diary to generate new statements in their name.

Beyond privacy, there is the issue of dignity and faithfully representing the person. Many cultures and ethical systems uphold the idea of respecting the dead – for example, not defaming them, and handling their memory with care (UN General Assembly 2024; Laqueur 2015; Metcalf and Huntington 2005). A griefbot, however, could potentially misrepresent the person's character or wishes. Because AI will inevitably generate novel answers, the bot will say things the real person never said. If those utterances are out of character or inaccurate, it could be seen as a violation of the deceased's "truth". In the best case, a well-trained bot stays very true to the person's known personality; but even then, it is going to be an approximation. In the worst case, a poorly constrained AI might produce offensive or false statements under the deceased's name. This possibility has led to warnings about griefbots' exploitability "to fulfill fantasies or used in ways that degrade the deceased" (Lim 2024). For instance, one might prompt a bot to say things the real person would never have consented to – effectively making the deceased perform digital "puppetry" against their will (Lim 2024). Such misuse clearly offends human dignity and has prompted calls for standards to protect the dead (UN General Assembly 2024). Practically, this could mean requiring that AI avatars only say things drawn from the person's actual recorded words, or giving families veto power over a bot's output. At a minimum, transparency is crucial: others interacting with the bot (or seeing its content) should know it's a simulation labeled as such, to avoid any misunderstanding that the real person said or did those things (indeed, following the rise and fall of Project December, OpenAI's own policy now demands such labeling for AI that "simulates" private individuals) (Hollanek and Nowaczyk-Basińska 2024).

The presence of digital ghosts might change how people relate to death and mourning. Culturally, reactions to this technology differ. In communities with traditions of ongoing communion with ancestors (such as parts of East Asia or among those who celebrate Día de los Muertos), an AI ghost may feel like a natural extension of remembering the dead (Klugman 2024). In more secular Western contexts that draw a firmer line between life and death, many find the idea "creepy" or inherently uncanny (Klugman 2024). There's also concern for certain groups: children, for example, are in a psychological stage where death's finality can be hard to grasp. If a child can still video-chat with their mom's avatar, will they struggle to accept that their mom is truly gone? Young children presented with an AI parent might insist the parent is still alive, possibly leading to confusion or developmental issues. Even among adults, there could be guilt or social pressure dynamics ("Why are you not talking to Dad's bot – don't you miss him?" or conversely "It's disrespectful that you keep talking to that thing"). And consider those who did not have a good relationship with the deceased: encountering that person's AI ghost unexpectedly (say a friend tries to "surprise" you with it, or a company uses it in marketing) could be traumatic (Klugman 2024). The technology opens new possibilities for retraumatization (imagine a domestic abuse survivor running into an AI of their abuser) if not carefully managed. Thus, the ethical deployment of griefbots must consider who uses them and under what circumstances, perhaps restricting use to those mature enough to handle it.

Another major concern is the profit motive behind many of these services. The "digital death market" is poised to grow, and companies see opportunity in offering to "immortalize" people or to "reunite" families with lost loved ones – for a price (The Guardian 2016; Grieshaber and Hadero 2024). This raises questions: Are griefbots being developed for the benefit of mourners, or to cash in on vulnerability? If it's the latter, there is a concrete risk of exploitative practices. For example, a company might create an AI ghost of a person (using whatever data can be scraped or purchased) without anyone asking, and then market it to the family – "We have your late grandfather's avatar ready, would you like to talk to him? Only $9.99 per month". This scenario is not far-fetched; Klugman has likened it to an amusement park snapping

your photo on a ride and then trying to sell it to you at the exit – but here the "ride" is your loved one's death (Klugman 2024). The ethical issue is whether we should allow a free-market, free-for-all in this sensitive arena. Unregulated, it could lead to abuses of data and trust, such as AI griefbots subtly pushing products, or even outright fraud. Notably, AI deepfakes have already been used in scams (impersonating voices to trick relatives out of money), and while those usually involve living persons, one can't rule out similar tactics around the deceased (Klugman 2024). The intersection of grief and money is always delicate (consider the criticism of predatory funeral industry practices in the U.S.) (Kopp and Kemp 2007a; 2007b; Mitford 1998), and introducing AI avatars could become another way to take advantage of people not thinking critically due to emotional distress. Guidelines or regulations can help prevent manipulative commercialization of grief.

## A taxonomy of digital afterlife technologies

To rigorously analyze these ethical and legal challenges, we also need a clear map of the technological landscape. Although research on AI-mediated digital afterlife services is still emerging, it is clear that these systems vary widely in form, purpose, and ethical complexity. Recent work has attempted to classify them in different ways. Hurshman and colleagues argue that "digital duplicates" (including digital ghosts) should be treated as context-specific tools, whose risks and safeguards depend on whether they are used for education, memorialization, advocacy, entertainment, or medical decision-making (Hurshman et al. 2025). A complementary approach focuses on the stakeholders involved. Hollanek and Nowaczyk-Basińska distinguish the data donor, the data recipient, and the service interactant; a structure that highlights questions of consent and responsibility, including the need for both premortem authorization and informed user consent (Hollanek and Nowaczyk-Basińska 2024).

However, these approaches remain incomplete on their own. Stakeholder and context classifications identify who is involved and why a system is used, but they do not fully capture the structural variation in how these systems are built: their data sources, fidelity, interactivity, governance, or behavioral agency. To address this, we propose a multidimensional taxonomy that maps the operative features shaping the ethical and legal profile of digital resurrection technologies.

### Timing

The first criterion is the timing of creation: whether the digital persona is developed before or after the subject's death. This distinction is foundational, as pre-mortem avatars typically reflect a form of intentional legacy planning, whereas post-mortem creations raise questions about the legitimacy of reanimation and the ethics of reconstructing a person without their input.

## Consent

Closely linked is the question of consent. Digital personas may be created with the explicit authorization of the individual during life, with surrogate consent from family members or legal proxies, or without any form of consent at all. Each of these conditions carries distinct ethical implications, especially in terms of dignity, posthumous privacy, and autonomy.

## Source(s) of data

The source of data used to train or construct the avatar further differentiates these systems. Some are based on material deliberately provided by the subject, such as interviews, written responses, or recorded media. Others are generated posthumously by extracting digital traces, such as emails, chat logs, or social media posts, often without clear permission. In some cases, the data is compiled or uploaded by third parties, typically family members or platform operators.

## Interaction

The mode of interaction and interactivity refers both to the medium through which users engage with the avatar – text, audio, video, or multimodal interfaces – and to the degree of dialogic responsiveness. Some systems are fully interactive, capable of adaptive conversation and responsive behavior. Others are semi-interactive, offering limited branching responses or scripted dialogue, while still others are non-interactive, functioning as passive representations or pre-recorded messages.

## Fidelity and disclosure

Equally important is the dimension of fidelity and disclosure. Fidelity refers to how closely the system emulates the subject's tone, language, behavior, or worldview. High-fidelity systems may be trained on extensive personal data and offer convincing likenesses, while lower-fidelity versions may rely on generic templates or sparse inputs. Disclosure concerns whether the system is clearly identified as artificial. Ethically robust implementations may include disclaimers, visual indicators, or conversational cues to clarify that the user is interacting with a simulation, not a sentient or conscious being.

## Purpose

The primary purpose of the system also plays a defining role. Digital afterlife technologies may be used for grief support, designed to maintain a form of contact with the deceased (Sri Takshara and Bhuvaneswari 2025; Riggs 2025; Obadia 2025); for legacy preservation, where individuals proactively record memories or stories (Lei et al. 2025); for entertainment or public engagement, especially in the case of celebrity recreations (Aboulnasr and Song 2025; Orita 2022); for education or historical simulation; or for instrumental applications such as legal testimony, proxy decision-making (Earp et al. 2024), or therapeutic scenarios.

### Audience and access

This intersects with the intended audience and access model. Some systems are designed strictly for private use – typically within families or among close friends – while others might be publicly accessible, used by fan communities or institutional stakeholders.

### Governance and ownership

The governance and ownership of the avatar includes both legal control and practical authority over the system. It matters whether the digital persona is owned or managed by the family, by a commercial platform, or by a public institution. This includes the ability to revise, delete, restrict, or even monetize the simulation, as well as the right to "retire" it at the discretion of relevant stakeholders.

### Autonomy and behavioral agency

Finally, the dimension of autonomy and behavioral agency addresses whether the avatar is simply reactive – responding only to user prompts – or exhibits a higher degree of initiative, such as initiating conversations, simulating memory, or evolving over time. Systems with greater behavioral agency may tend to evoke stronger emotional responses and may lead users to anthropomorphize the simulation more intensely, raising concerns about dependency, blurred boundaries, and emotional entanglement.

Together, these dimensions define the operative space of digital resurrection technologies. They provide a structured foundation for analyzing the ethical challenges posed by specific cases, from consensual legacy avatars to unauthorized griefbots or celebrity simulations. This taxonomy also enables differentiation between systems that may appear similar on the surface but diverge significantly in terms of moral legitimacy, emotional impact, and regulatory risk.

The table below applies the nine-dimensional taxonomy discussed above to a set of empirical archetypes previously discussed – namely, legacy avatars, private griefbots, celebrity simulations, and institutional avatars. These configurations represent the most publicly visible and operational forms of AI-mediated digital afterlife services currently in circulation.

| Type → <br> Category ↓ | Legacy Avatars | Private Griefbots | Celebrity Simulations | Institutional Avatars |
|---|---|---|---|---|
| *Timing of Creation* | Pre-mortem | Post-mortem | Post-mortem | Pre- or post-mortem |
| *Consent* | Explicit (user-authored) | None or surrogate | Rare or unclear | Explicit or surrogate |
| *Data Source* | Direct input (e.g. interviews, Q&A) | Digital traces (texts, chats, social media) | Public media archives | Curated archival or testimonial material |
| *Interaction & Interactivity* | Audio/video; semi-interactive | Text/audio; fully interactive | Video/audio/chat; sometimes interactive | Video/avatar; semi-interactive |
| *Fidelity & Disclosure* | Medium–high fidelity; clearly disclosed | Medium–high fidelity; disclosure varies | Medium fidelity; often not disclosed clearly | Medium fidelity; disclosed as artificial |
| *Primary Purpose* | Legacy preservation | Grief support | Entertainment or tribute | Education, legal or therapeutic use |
| *Audience / Access* | Private (family or close circle) | Private (individual use) | Public | Institutional / restricted public |
| *Governance / Ownership* | Controlled by user or family | User or informal developer control | Corporate or fan-operated | Institutional or estate-managed |
| *Autonomy / Behavioral Agency* | Low–moderate (scripted or fixed) | Moderate–high (adaptive conversation) | Moderate (can initiate interactions) | Low–moderate (structured behavior) |

Table 1. Archetypal configurations of digital afterlife technologies. This table classifies four empirically grounded types of digital afterlife systems using a nine-dimensional taxonomy. Each archetype reflects a distinctive configuration of timing, consent, data source, interactivity, fidelity, purpose, access, governance, and behavioral autonomy, offering a basis for ethical comparison and regulatory foresight.

This taxonomy highlights that digital ghosts are not a single phenomenon but a plethora of use cases shaped by different configurations dimensions – and therefore, by different configurations of ethical issues. For example, both HereAfter AI and Project December involve emotionally significant simulations, yet the former is designed as a consensual legacy tool, while the latter enabled spontaneous post-mortem dialogue without prior authorization. Likewise, celebrity deepfakes and courtroom avatars might both be video-based, but they serve radically different functions: one commercial, the other built for evidence in court.

Yet the taxonomy does more than describe the present: it also opens space for imagining the near future. By recombining values across the nine dimensions, we can forecast other plausible archetypes that have not yet reached mainstream use but are technologically and socially plausible. Some of these prospective forms may be benign or therapeutic, such as simulacra used for trauma therapy or AI agents designed to execute a person's ethical wishes after death. Others raise more troubling possibilities.

Among the speculative forms that merit attention are communal memorial bots sustained by crowdsourced memories (Brubaker 2013; Kozlovski and Makhortykh 2025); therapeutic confrontations with abusive or estranged figures reconstructed without consent (while no published study endorses reconstructing an abusive figure, the psychological risks discussed in literature (Krueger and Osler 2022; Voinea et al. 2025) make the extrapolation plausible and well grounded); doppelgangers or "living ghosts" developed pre-mortem for legacy rehearsal or self-dialogue (Register et al. 2025; Schwitzgebel et al. 2024); and AI agents built to simulate decision-making for legal or institutional continuity (Earp et al. 2024). At the far end of the ethical spectrum lie explicitly malicious uses, including impersonation of the dead to deceive survivors, posthumous harassment or defamation via synthetic content, or politically motivated resurrections of martyrs for ideological propaganda. These latter scenarios signal urgent governance challenges as AI capabilities scale and barriers to entry lower.

# Ethical considerations

With these dimensions in place, we can now examine when and how digital ghosts become ethically acceptable, and what conditions must be met to prevent harm. If the taxonomy above defines the operational space of digital afterlife technologies, it also allows us to articulate how an ethically designed digital ghost should be developed. An "ethical digital ghost" is a configuration of choices across the nine dimensions that collectively minimize harm, respect the deceased, and support the well-being of the living. Based on the dimensions outlined above, several core features emerge.

First, an ethical digital ghost should be rooted in premortem intent. Its creation should follow from the explicit, informed decision of the person during life, rather than post-mortem reconstruction. This establishes a foundation of autonomy: the subject knowingly authorizes the form and scope of their digital presence, much as they might prepare an advance directive or a will. Furthermore, meaningful consent should govern not only the data donor but also those who interact with the system. Ethical implementations should ensure that the deceased authorized the use of their likeness and data, and that surviving users consciously opt in, with a clear understanding of what the system is and is not. No avatar should be created or deployed without the affirmative permission of every party implicated. Moreover, the system should draw strictly on user-provided and consented data, avoiding invasive mining of private archives or scraping of digital traces never intended for this purpose. Data stewardship should be transparent: users should know what data was used, how it was processed, and who controls it. Arguably, an ethical digital ghost should also maintain strict disclosure and calibrated fidelity. The system must make its artificial nature unmistakable, through both visual and conversational signals, preventing users from confusing simulation with survival. High fidelity may be permissible, but only when paired with unambiguous reminders that the avatar is a representation, not a continuation of consciousness. Fidelity should support remembrance, not impersonation. Further, its purpose should be narrowly tailored to humane and proportionate aims: legacy preservation or grief support, not commercial exploitation, political persuasion, or legal substitution. Ethical digital ghosts should be tools of memory, not instruments of influence. It should be considered that uses such as that of Chris Pelkey's impact statement pose serious fidelity issues: it is arguable that the experience of dying could have significantly altered Pelkey's worldview; however, it is structurally impossible to include the subjective element of that experience in the dataset of Pelkey's digital ghost. Therefore, as its fidelity is questionable, and fidelity is a crucial factor for such a use case, that use case should be excluded. Furthermore, access and audience should be limited. These systems should be designed primarily for private, voluntary use, rather than public display or broadcast. A loved one's likeness should not be turned into a spectacle, a meme, or a political symbol without clear, explicit permission. Also, the governance and ownership of the avatar should lie with the person's designated heirs or estate, not a commercial platform. Families or executors should have the authority to revise, pause, or permanently retire the system. In this sense, the digital ghost functions as a form of residual personhood (Spitale 2015, 122), a posthumous extension of the self whose moral stewardship properly belongs to those entrusted with the person's memory and affairs. Commercial actors would serve only as custodians, bound by fiduciary-like duties to protect the dignity of the deceased and the interests of users. Finally, an ethically designed digital ghost should be endowed with minimal behavioral agency. It should not simulate new memories, or evolve beyond the boundaries set by the deceased. Its function is archivistic rather than generative: to preserve, not to reinvent. Limiting agency reduces the risk of emotional over-attachment, distortion of identity, or the uncanny sense that the deceased is "acting" in ways they never would have chosen.

We argue that these design principles illustrate that ethical digital ghosts are defined by their alignment with the principles of autonomy, transparency, dignity, and emotional safety. The goal should not be to create a digital substitute for the deceased, but to provide a carefully bounded, consensual medium through which the living can remember without being deceived, comforted without being manipulated, and connected without losing sight of loss.

## Sources of normativity

These ethical judgments rest on deeper normative commitments, which we now make explicit. The normative foundation for defining what an ethical digital ghost should look like emerges from three interlocking domains of value that converge in the phenomenon itself. First, the nine-dimensional taxonomy outlined above is not merely descriptive: each dimension highlights a morally salient feature of digital afterlife systems. Timing and consent encode respect for autonomy; data source and governance index privacy, control, and stewardship; fidelity and disclosure speak to truthfulness and the avoidance of deception; interaction modality and behavioral agency track risks of emotional harm, dependency, or distortion. The taxonomy therefore identifies the structural sites where ethical evaluation is required and the levers through which ethical practice can be enacted. In this sense, the design of an ethical digital ghost is the value-aligned configuration of these dimensions.

Second, these value commitments are reinforced by the convergent principles that permeate the existing literature on griefbots, posthumous privacy, and digital personhood. Across bioethics, data protection scholarship, and grief psychology, four core concerns consistently reappear: respect for autonomy, protection of dignity, safeguarding of privacy, and prevention of harm (Hollanek and Nowaczyk-Basińska 2024; Hurshman et al. 2025; Lindemann 2022; Lim 2024). When we call for premortem consent, transparent disclosure, limits on behavioral agency, and protections against commercial exploitation, we are responding to these same core values. Thus, the ethical design of digital ghosts is anchored in a broader cross-disciplinary consensus about what is at stake.

Third, the normative horizon is shaped by long-standing social norms governing the treatment of the dead. Across cultures, the deceased are afforded symbolic dignity (Metcalf and Huntington 2005), their memory is protected from distortion (Laqueur 2015; Verdery 1999), and their remains (including, for some scholars, digital or informational remains) are not to be appropriated without permission (UN General Assembly 2024; Birnhack and Morse 2022; Richardson 2015). The emergence of digital ghosts does not erase these expectations; it tests them. The moral unease triggered by unauthorized resurrections, manipulative uses, or "haunted" interactions reflects inherited cultural intuitions about reverence, truthfulness, and care for the bereaved. Ethical digital ghosts therefore draw not only on contemporary tech-ethics but on deep cultural commitments that predate these technologies by centuries.

# Conclusion

The rise of deadbots and digital ghosts forces us to confront fundamental questions about how we remember our dead. These AI replicas sit in a delicate space between cherished memory and possible deception, between extending love and prolonging pain. Digital ghosts can become the blind spot between memory and trickery, a prolonged mourning disguised as dialogue. We must ask: when does holding on through technology help us, and when does it hold us back? Is it an act of love to keep a part of someone "alive" in bits and code – or do we risk cheating ourselves of real closure, "technologically assisted" in our difficulty to let go?

From an ethical standpoint, arguably intent and transparency make a substantial difference. Creating or engaging with a simulacrum of a loved one knowing fully that it is an artificial comfort could be viewed as a personal choice – a modern form of remembrance – provided it's done with eyes wide open, and harms no one. If it helps soothe the pain of grief in the short term, some would argue it can be compassionate and even morally right (Wright 2024). However, doing the same thing without honesty – either by fooling someone into thinking the AI is more than what it is, or by self-delusion – edges into dangerous territory. Similarly, from a societal perspective, allowing people to have these tools might be beneficial (given how many already talk to gravestones or pray to deceased ancestors, an AI might simply reply back). Yet we also have a duty to guard against the potential abuses – exploitative market logics, privacy invasions, and distortions of a person's legacy.

Legally and policy-wise, the emerging consensus is that we cannot leave this entirely to the free market, especially given the fragmented governance structures that currently characterize digital remains, where platform policies, national laws, and cultural norms collide without clear hierarchy or oversight. As Klugman noted, we have to decide whether to leave AI ghosts to the marketplace, to ban them, or to regulate them (Klugman 2024). Given the stakes, outright laissez-faire seems risky; a ban on all such technology may be neither feasible nor necessary (especially if it truly helps some people); thus, sensible regulation and guidelines are arguably the most prudent path. This includes setting standards for consent, data use, age restrictions, and marketing practices – essentially placing ethical guardrails on an industry that uniquely fuses our age-old "quest for immortality" with modern capitalism (Grieshaber and Hadero 2024). We should ensure that human values of dignity, privacy, and compassion aren't overshadowed by profit motives.

Ultimately, digital ghosts compel us to reckon with how technology mediates one of the most human experiences: grieving. They challenge us to balance the fidelity to the truth of the person who died – not turning them into something they were not – with empathy for the living who mourn. Society's comfort with these tools may evolve (just as live-streamed funerals, once seen as odd, became normal during the pandemic) (Klugman 2024), but we should consciously shape this evolution rather than be blindsided by it.

The era of AI "afterlives" is here, and it falls upon us to ensure this technology is used in a way that supports memory without becoming an imposture, and helps heal without betraying the dignity of those we love and lose. Griefbots should be tools that assist in grief, not a permanent substitute for the departed (Morris and Brubaker 2025).

## Author contributions

GS: Conceptualization; Critical analysis; Writing – original draft; Writing – review & editing.

FG: Validation; Writing – review & editing.

# Acknowledgments

GS thanks Tiglia Panevinos, steadfast companion and quiet witness, for reminding us that the line between presence and absence can be tender, gentle, and very much alive.